\documentclass[floatfix,showpacs,amsmath,amssymb,aip,jmp,preprint]{revtex4-1}
\usepackage[USenglish]{babel}
\usepackage{overpic}
\usepackage[urlcolor=blue,colorlinks=true,linkcolor=blue,pdfstartview={FitH}]{hyperref} 

\newcommand{\ket}[1]{\ensuremath{|#1\rangle}}
\newcommand{\bra}[1]{\ensuremath{\langle#1|}}

\newcommand{\eg}{\emph{e.g.}}
\newcommand{\ie}{\emph{i.e.}}
\newcommand{\etal}{\emph{et al.}}

\newcommand{\smal}{\scriptscriptstyle}
\newcommand{\mc}{\mathcal}
\newcommand{\mb}{\mathbb}
\newcommand{\ot}{\otimes}
\newcommand{\op}{\oplus}

\newcommand{\lm}{\lambda}
\newcommand{\al}{\alpha}

\newcommand{\idQ}{\mb{I}_{Q}}
\newcommand{\idE}{\mb{I}_{E}}

\begin{document}

\title{Riccati equation and the problem of decoherence II: Symmetry and the solution of the Riccati equation}
\date{28 February 2010}

\author{Bart{\l}omiej Gardas}
\email{bartek.gardas@gmail.com}

\affiliation{Institute of Theoretical and Applied Informatics, Polish Academy
of Sciences, Ba{\l}tycka 5, 44-100 Gliwice, Poland}

\begin{abstract}
In this paper we revisit the problem of decoherence
applying the block operator matrices analysis. Riccati
algebraic equation associated with the Hamiltonian
describing the process of decoherence is studied.
We prove that if the environment responsible for 
decoherence is invariant with respect to the antylinear
transformation then the antylinear operator solves
Riccati equation in question. We also argue that this
solution leads to neither linear nor antilinear operator
similarity matrix. This fact deprives us the standard
procedure for solving linear differential equation (\eg, Schr\"odinger equation).
Furthermore, the explicit solution of the Riccati equation is
found for the case where the environment operators commute with
each other. We discuss the connection between our results and the standard
description of decoherence (one that uses the Kraus representation).
We show that reduced dynamics we obtain does not have the Kraus
representation if the initial correlations between the system 
and its environment are present.
However, for any initial state of the system (even when the 
correlations occur) reduced dynamics can be written in a manageable way.  
\end{abstract}
\maketitle



\section{Introduction} \label{sec:intro}
 
Recently, the connection between a problem of decoherence and the Riccati
operator equation was established (for details see~\cite{gardas} and Ref.
therein). Moreover, it was shown that a wide class of a time-dependent quantum
systems, precisely the ones that describe a qubit $Q$ interacting with the
environment $E$ and defined by the following Hamiltonian

\begin{equation}  
 \label{hamil}
   H_{QE}(t,\beta) = H_{Q}(t,\beta)\ot\mb{I}_{E}
                    +\mb{I}_{Q}\ot H_{E}+f(\sigma_3)\otimes V, 
\end{equation}
where the Hamiltonian $H_{Q}(t,\beta)$ of the qubit alone is give by

\begin{equation}
  \label{open}
H_Q(t,\beta)= \beta\sigma_3
            +\alpha\left(\sigma_1\cos(\omega t)+\sigma_2\sin(\omega t)\right),
\end{equation}
where $\alpha$, $\beta\in\mb{R}$, can be \emph{effectively} simplified to the
time-independent problem. Namely the one governed by the Hamiltonian
$H_{QE}(0,\beta)\equiv H_{QE}$. The connection between the reduced dynamics
~\cite{alicki} of those models can be summarized in the following formula

\begin{equation}
 \label{reduce}
 \bar{\rho}_t = V_t\rho_t(\bar{\beta})V_t^{\dagger},
\end{equation} 
where $\bar{\beta}:=\beta-\omega /2$ plays the role of effective parameter 
and $V_t:=\mbox{diag}(e^{-i\omega t/2},e^{i\omega t/2})$ is the unitary (similarity)
transformation. Here, $\bar{\rho}_t$ is the solution (reduced dynamics) of the system
identified with the time-dependent Hamiltonian $H_{QE}(t,\beta)$ and $\rho_t(\beta)$ 
representing the reduced dynamics of the model described by the Hamiltonian $H_{QE}$. 
An explicit dependence $\bar{\rho}_t$ of $\beta$ was omitted. The meaning of the
symbols we used in the equations~(\ref{hamil}) and~(\ref{open}) is usual. Furthermore, 
it was found that this time-independent problem can be solved in two ways. One could 
either resolve for $X$ the Riccati operator equation:

\begin{equation}\label{riccati}
\alpha X^2+X(H_{+}+\beta)-(H_{-}-\beta)X-\alpha=0,
\end{equation}
where $H_{\pm}:=H_{E}\pm V$, or solve the following Schr\"{o}dinger equation
(we work with the units $\hbar=1$)

\begin{equation}\label{schrodinger}
i\dot{\Psi}_{t}=H_t\Psi_t, \quad 
H_t=
\left[ 
\begin{array}{cc}
 H_{E} & z_t^{\ast}V_{\beta}   \\
 z_tV_{\beta} & H_{E} 
\end{array}
\right],
\end{equation}
with initial condition $\Psi_0 \equiv \Psi$. In Eq.~(\ref{schrodinger})
$\Psi_t=[\psi_t,\phi_t]^T\in\mc{H}\op\mc{H}$, $z_{t}=e^{-i2\alpha t}$ and
$V_{\beta}=V+\beta\mb{I}_{E}$. In the description above it was assumed 
that $\mc{H}$ is a separable Hilbert space (possibly infinite-dimensional) and
$H_{E}$ and $V$ are the Hermitian operators acting on it. One could recall that
in the current paper we assume that the Riccati equation $R_{H}[X]=0$, where

\begin{equation}
  R_{H}[X] = XBX+XA-CX-B^{\dagger},
\end{equation}
is associated with the following Hamiltonian

\begin{equation}
H=
\begin{bmatrix}
A & B\\
B^{\dagger} & C
\end{bmatrix}.
\end{equation}
In turn if the solution of the equation $R_{H}[X]=0$ exists, it may
be used to diagonalize operator matrix $H$. Following equality holds true

\begin{equation}\label{diag1}
S_{X}^{-1}HS_{X}=
\begin{bmatrix}
A+BX & 0 \\
0 & A-(XB)^{\dagger}
\end{bmatrix},
\end{equation}
where $S_{X}^{-1}$ stands for the inversed operator matrix to 

\begin{equation}
 \label{sx}
S_{X} = 
\begin{bmatrix}
\idE & -X^{\dagger} \\
X &  \idE
\end{bmatrix}.
\end{equation}  
The Riccati Eq.~(\ref{riccati}) is associated with the Hamiltonian $H_{EQ}$.
On the other hand the Riccati equation related to the Hamiltonian $H_t$ reads

\begin{equation}
 \label{triccati}
 X(z_t^{\ast}V_{\beta})X+XH_{E}-H_{E}X-z_tV_{\beta}=0.
\end{equation}
One should keep in mind that the given operator $A$ acting on the space $\mb{C}^{2}\ot\mc{H}$
may be thought of as a block operator matrix, since the following isomorphism holds
$\mb{C}^{2}\ot\mc{H}=\mc{H}\oplus\mc{H}$. 
 
It is worth mentioning that in spite of the fact that connection between operator matrices
$H_t$ and $H_{QE}$ is well defined and the solution of the Eq.~(\ref{triccati}) can be easily
obtained (indeed, it is given by $X_t=z_t\mb{I}_{E}$ ) the solution of the Eq.~(\ref{riccati})
is still missing. Please note that although $X_t$ is known it cannot be effectively used to
resolve Eq.~(\ref{schrodinger}) because of its explicit time dependence. Nevertheless it 
allows us to diagonalize the Hamiltonian $H_t$, namely

\begin{equation}
  \label{hd}
  S_{z_t}^{\dagger}H_tS_{z_t}
              = 
\left[ 
\begin{array}{cc}
 H_{+}+\beta\mb{I}_{E} & 0   \\
 0 & H_{-}-\beta\mb{I}_{E} 
\end{array}
\right] 
\equiv  H^{d}, 
\end{equation} 
where the unitary matrix $S_{z_t}$ is defined as 

\begin{equation}\label{sz}
S_{z_t}=
\frac{1}{\sqrt{2}}
\left[ 
   \begin{array}{cc}
         \mb{I}_{E} & -z_t^{\ast}\mb{I}_{E} \\
          z_t\mb{I}_{E} & \mb{I}_{E} 
   \end{array}
\right]
=\frac{1}{\sqrt{2}}
\left(
   \begin{array}{cc}
      1 & -z_t^{\ast}   \\
      z_t & 1
\end{array}
\right)\ot\mb{I}_{E}.
\end{equation}
We want to emphasize that the Eq.~(\ref{triccati}) and its slight modification 
 
\begin{equation}
 z_t^{\ast}XV_{\beta}X+XH_{E}-H_{E}X-z_tV_{\beta}=0,
\end{equation} 
used by the author in previous manuscript on the subject~\cite{gardas}, are equivalent only
if the solution is assumed to be a linear operator. This seems to be justified, especially
if one expects $X$ to represent an observable. However, it does not need to be true. Thus in
this manuscript we will not restrict the analysis to the linear operators only. 
 
The purpose of this manuscript is two fold. Firstly, we show that if the environment 
is (\ie, the operators $H_E$ and $V$ are) invariant under antylinear transformation 
(Sec.~\ref{sym}), then this antylinear operator is the solution of the Riccati equation
~(\ref{triccati}). We also indicate that mentioned symmetry may be used to diagonalize
the Hamiltonian related with Eq.~(\ref{riccati}), although it does not solve this equation. 
Next, in Sec.~(\ref{problem}) we argue that the solution we obtained can not be applied to
solve Eq.~(\ref{schrodinger}). The problem occurs because standard methods provided by the
theory of the differential equations demand that the operator is not antylinear. The reduced
dynamics of the system under consideration is given in Sec.~(\ref{dynamics}).
 
Secondly, in Section~(\ref{solution}) we study the case when $[H_E,V]=0$ and we give
the exact solution of the Riccati equation~(\ref{riccati}) for this situation. This 
is direct generalization of the recently found solution for the particular operator
defining the spin-environment. Finally, in Sec.~(\ref{kraus}) we compare our method
with the standard approach based on the operator sum representation. We also discuss
the possibility of obtaining an operator (or Kraus) sum representation in the case 
when the correlations between the system and its environment are present initially.

\section{Symmetry and the solution to the Riccati equation}
  \label{sym}

Let $\tau_1$ and $\tau_2$ be an antilinear symmetry for $H_{E}$ and $V$,
such that $\tau_i^2=\mb{I}_{E}$, (or $\tau_i=\tau_i^{-1}$) for $i=1,2$, respectively. By 
definition, it means that $[H_{E},\tau_1]=0$ and $[V,\tau_2]=0$. Since the symmetry
operator is an involution \ie, $\tau^2=\idE$, thus for a given operator $A$ the condition 
$[\tau, A]=0$ is equivalent to the equality $\tau A\tau^{-1} =A$. Operators that fulfill
last equality are invariant under the action of $\tau$ and are called $\tau-$symmetric. 
Furthermore, the statement that an operator $\tau$ acting on $\mc{H}$ is an antilinear has
the following meaning

\begin{equation}
 \tau(a\psi+b\phi)= a^{\ast}\tau\psi+b^{\ast}\tau\phi,
\end{equation}
for every $\psi$, $\phi\in\mc{H}$ and $a$, $b\in\mb{C}$. We wish to
emphasize that in the finite-dimensional case the existence of the aforementioned symmetry
is ensured by the theorem of Ali Mostafazadeh~\cite{ali1} which states that every 
diagonalazable pseudo-Hermitian (in particular the Hermitian) operator $H$ with
the discrete spectrum has an antilinear and anti-hermitian symmetry $\tau$. Moreover,
this symmetry is an involution, \ie, $\tau^2=\idE$, if $H$ is Hermitian operator, \ie,
$H=H^{\dagger}$.  The proof of this theorem provides the explicit construction
of $\tau$. Let us additionally assume that $\tau_1=\tau_2:=\tau$, which means that the
operators $H_{E}$ and $V$ posses a \emph{common} symmetry $\tau$.

If one allow  $X$ to be an antilinear operator, then equation~(\ref{triccati}),
can be rewritten in the following form

\begin{equation}
 R_{H_t}[X] = z_t\left(XV_{\beta}X-V_{\beta}\right)+[X,H_{E}].
\end{equation}
From this equation we can readily see that $R[\tau]=0$, \ie, the symmetry
generator $\tau$ of the $E$ system is the solution we were seeking for.
In particular, if the Hamiltonians $H_{E}$ and $V$ are 
(or roughly speaking the $E$ system is) $PT-$symmetric (or $T$-symmetrix) where $P$ and $T$
stand for the parity and the time-reversal operators, respectively, then $X=PT$ ($X=T$),
\ie, the $PT$ ($T$) operator solves Eq.~(\ref{triccati}). This is very interesting and rather
unexpected result that the famous $PT$ ($T$) symmetry is the solution of the Riccati equation we study.

For instance it can be easily proven that if both $H_{E}$ and $V$ are symmetric operators, that is
to say $H_{E}^{T}=H_{E}$  and $V^{T}=V$, where by ``$T$'' we denoted the operation of transposition,
then it implies that $R_{H_t}[K]=0$, where $K$ is complex conjugate operator (see below). In order to
do that let us define 

\begin{equation}\label{ccp}
 K\psi = \psi^{\ast}, \quad \psi\in\mc{H},
\end{equation}
where by ``$\ast$'' stands for the standard complex conjugation operation.
The operator above possesses the following properties: 

\begin{enumerate}

 \item[$a)$] $K=K^{\dagger}$ (\ie, it is  Hermitian operator),
 \item[$b)$] $KK^{\dagger}=\mb{I}_{E}$ (\ie, it is  unitary operator),
 \item[$c)$] $K^2=\mb{I}_{E}$ (\ie, it is an involution).
\end{enumerate}
Moreover, $K$ is  antilinear operator. The listed properties follow
immediately from the definition~(\ref{ccp}). Note that for any Hermitian
operator (matrix) $A$ the condition $A^{T}=A$  means that
$A$ is $K-$symmetric, \ie, $[A,K]=0$. To see this observe that $K$
transforms any operator $A$ in accord with the following rule:

\begin{equation}\label{tras}
 KAK^{\dagger}= A^{\ast}.
\end{equation}
In addition, if one assumes that $A$ is Hermitian, \ie, $A=A^{\dagger}$
and uses properties $a)$ then from the equation above we have

\begin{equation}\label{transpose}
  KAK=A^{T},
\end{equation}
where we used the fact that $A^{\dagger}=(A^{T})^{\ast}$.
Clearly, for the symmetric matrix $KAK=A$ or 
$[A,K]=0$. Other way to see that $R_{H_t}[K]=0$ is
to rewrite $R[K]$ as

\begin{equation}\label{rk}
R_{H_t}[K]=z_t(V_{\beta}^{T}-V_{\beta})+(H_{E}^{T}-H_{E})K.
\end{equation} 
Evidently, for the symmetric operators the right side of the Eq.~(\ref{rk}) vanishes.  

At the end of this section we show how to diagonalize the operator $H_{QE}$
using the symmetry $\tau$. Firstly, note that if one introduces the
unitary operator $U$ in a way that

\begin{equation}\label{udiag}
U=
\frac{1}{\sqrt{2}}
\begin{bmatrix}
\idE & i\idE \\
i\idE & \idE
\end{bmatrix}
=\frac{1}{\sqrt{2}}
\begin{pmatrix}
 1 & i \\
 i & 1
\end{pmatrix}
\otimes\idE,
\end{equation}
then $U^{\dagger}H_{QE}U=\bar{H}$ and
\begin{equation}
\bar{H}
=
\begin{bmatrix}\label{hbar}
 H_{R} & V_{\alpha\beta} \\
 V_{\alpha\beta}^{\dagger} & H_{R}
\end{bmatrix},
\end{equation}
where  $V_{\alpha\beta}:=\alpha\idE-iV_{\beta}$. Since $\tau V_{\alpha\beta}\tau=
V_{\alpha\beta}^{\dagger}$, thus from Eq.~(\ref{hbar}) we see that $R_{\bar{H}}[\tau]=0$,
 \ie, the symmetry $\tau$ is the solution of the Riccati equation associated with the 
Hamiltonian~(\ref{hbar}). Therefore, the matrix $\bar{S}_{\tau}:=US_{\tau}$ diagonalizes
the Hamiltonian $H_{QE}$. To be specific, the following equation holds true

\begin{equation}
\bar{S}_{\tau}^{\dagger}H_{QE}\bar{S}_{\tau}
=
\begin{bmatrix}\label{hdiag}
 H_{R}+V_{\alpha\beta}\tau & 0_{E} \\
 0_{E} & H_{R}-V_{\alpha\beta}^{\dagger}\tau
\end{bmatrix}.
\end{equation}
Keep in mind that the operator $V_{\alpha\beta}$ is \emph{not} Hermitian,
while the operator $V_{\alpha\beta}\tau$ is.
Interestingly, we diagonalized the block operator matrix $H_{QE}$ without directly   
resolving the Riccati algebraic equation associated with it. One may ask if this is 
possible in general. To be specific if there exists a matrix $T$ such that it transforms
a given operator matrix $A$ to another one $\bar{A}$ viz $\bar{A}=T^{-1}AT$
and the Riccati equation $R_{\bar{A}}[X]=0$ related to $\bar{A}$ has the solution 
$X_{\bar{A}}$ that is easy to be found. If that would be the case, than one may
construct similarity operator matrix, namely $S_{X_{\bar{A}}}T$ that diagonalizes $A$.
We saw that if $A=H_{QE}$, then $T=U$ (Eq.~(\ref{udiag})) and
$\bar{A}=\bar{H}$ (Eq.~(\ref{hbar})), thus the procedure is working. Therefore, this 
strategy is worth to be studied. However, we will not focus on this subject herein.

\section{The problem with the Schr\"{o}dinger equation}
 \label{problem}

The results of the paper~\cite{gardas} and the last section show
that $R_{H_t}[z_t]=0$ and $R_{H_t}[\tau]$, where $R_{H_t}[X]$ stands for
the left side of the Eq.~(\ref{triccati}). Note that, the second solution 
($X=\tau$) we obtained, in contrast to the first one ($X_t=z_t$) 
is time independent. Therefore, one may think that the second solution
has the advantage because it allows one to construct the operator matrix
$S_{\tau}$ that diagonalizes Hamiltonian~(\ref{schrodinger}) and do not
depend on time. Indeed this is the case however one serious drawback 
arises. To see this clearly, let us give an explicit form of the matrix 
$S_{\tau}$. In agreement with Eq.~(\ref{sx}) it takes the form
(compare this with Eq.~(\ref{sz}))

\begin{equation}\label{st}
S_{\tau}=
\frac{1}{\sqrt{2}}
\left[
\begin{array}{cc}
  \mb{I}_{E} & -\tau \\
  \tau &  \mb{I}_{E}
\end{array}
\right].
\end{equation}
Since the symmetry $\tau$ is an involution operator ($\tau^2=\mb{I}_{E}$), 
thus $S_{\tau}$ is the unitary matrix ($S_{\tau}S_{\tau}^{\dagger}=\mb{I}_{E})$.
We see that the similarity transformation~(\ref{st}) is neither linear nor 
antilinear operator. In particular, one may easily verify that
$S_{\tau}i=iS_{\tau}^{\dagger}$. All this difficulties are direct consequence
of the fact that the operator $\tau$ is antilinear. As a result, we cannot use the
standard procedure that allows us to solve linear, differential equation to 
resolve the Schr\"{o}dinger equation $i\ket{\dot{\Psi}_t}=H_t\ket{\Psi_t}$.
Indeed, because of the presence of the factor $i$ on the left side of the 
Schr\"{o}dinger equation we cannot apply the $\ket{\Phi_t}=S_{\tau}\ket{\Psi_t}$
transformation to reduce it to the following
form  $i\ket{\dot{\Phi}_t} = H_t^d\ket{\Phi_t}$, where $H_t^d$ stands for the
diagonal form of $H_t$. Another technique is needed to resolve this difficulties.
Unfortunately, at the present time there is none. Of course, the same problems 
occur when one tries to resolve the equation $i\ket{\dot{\Psi}_t}=H_{QE}\ket{\Psi_t}$
using the Eq.~(\ref{hdiag}).
 
Notice that according to procedure explained in the paper~\cite{gardas} the $H_t$
given by Eq.~(\ref{schrodinger}) has the following block-diagonalization
(compare with Eq.~(\ref{hd}))

\begin{equation}\label{hdt}
S_{\tau}^{\dagger}H_tS_{\tau}
=
\left[
\begin{array}{cc}
H_{E}+z_{t}^{\ast}V_{\beta}\tau & 0 \\
0 & H_{E}-z_{t}V_{\beta}\tau
\end{array}
\right]
\equiv
H_{t}^{d}.
\end{equation}
Comparing equations~(\ref{hd}) and~(\ref{hdt}) one can learn that the
solution $X_t=z_t$ leads to the time-independent diagonal form $H^d$
of the operator $H_t$, but the transformation matrix $S_{t}$ does depend 
on time. Opposite situation takes place in the case of the second solution 
$X=\tau$, \ie, diagonal form $H_{t}^{d}$ is time-dependent and
transformation matrix $S_{\tau}$ do not depend on time.

\section{Solution of the Riccati equation.}\label{solution}
Recently, the solution of the model specified by the Eq.~(\ref{hamil}) was given in
the case when $\mc{H}=\overset{N}{\underset{n=1}{\bigotimes}}\mb{C}^2$ and the
operator $H_{E}$, $V$ are defined as (for detailed discussion see~\cite{gardas2})

\begin{equation}\label{spin1}
  H_{E}= \sum_{n=1}^{N}\omega_n\sigma_3(n),\quad
  V= \sum_{n=1}^{N}g_n\sigma_3(n),
\end{equation}
where $\omega_n$ and $g_n$ are certain constants defining the frequencies and the coupling
constant of the spin-bath, respectively. For every $n\le N$ the operator $\sigma_3(n)$ is
understood as $\sigma_3(n)=I_2\ot...\ot\sigma_3\ot...\ot I_2$, where $I_2$ is a $2\times 2$
identity matrix and $\sigma_3$ is the usual Pauli matrix. Note that the 
operators above commute, \ie, $[H_{E},V]=0$. We will extend results obtained in 
paper~\cite{gardas2} to the arbitrary operators that commute with each other. 

If one assumes that the operators $H_{E}$ and $V$ commute, then 
they have common set of eigenvectors. Henceforward, we will
assume that the eigenvalues of $H_{E}$ and $V$ are all discrete and not degenerated.
Furthermore, the spectrum of a given operator $A$ will be denoted by $\sigma(A)$. 
As a summary, one can write the following eigenvalue problems for $H_{E}$ and $V$:

\begin{equation}\label{eigen}
 H_{E}\ket{\phi_n}=E_n\ket{\phi_n},\quad V\ket{\phi_n}=V_n\ket{\phi_n}. 
\end{equation}
where $E_n\in\sigma(H_{E})$ and $V_n\in\sigma(V)$. Here, the index $n$ goes through the set
of all integer numbers or through each subset of it. We wish to emphasize that the assumption of the
discreetness and no degeneration of the spectrum of the operators in question is not
crucial in our analysis. In fact, it can by easily overcome (however, we will not address 
this technical issue in the current manuscript).

Note, the solution $X$ of the Eq.~(\ref{riccati}) is a function of the operators $H_{\pm}$ 
defined in the Eq.~(\ref{riccati}). Since the operators $H_{E}$ and $V$ commute with 
each other, so are the operators $H_{\smal{\pm}}$. Therefore, $[X, H_{\smal{\pm}}]=0$ and 
the Riccati equation~(\ref{riccati}) can be simplified to the following form

\begin{equation}\label{sricc}
 \alpha X^{2}+2V_{\beta}X-\alpha = 0.
\end{equation}
One can observe that if $\alpha=0$ then $X_0=0_{E}$ is the
solution of the Eq.~(\ref{sricc}). No wonder since in this case matrix $H_{QE}$ is already in
the diagonal form. Yet, the operator $X_0$ may not be the only solution of the equation
$R_{H_{QE}}[X]=0$.

In order to obtain the solution for the case where $\alpha\not=0$ we will apply the
spectral theorem for Hermitian operators. From the Eq.~(\ref{sricc}) one can readily
see that $X=f(V_{\beta})$, where the function $f$ takes the form

\begin{equation}\label{fun}
 f(\lambda)= \frac{-\lambda+\sqrt{\lambda^2+\al^2}}{\al},
 \quad \lambda\in\sigma(V_{\beta}),
\end{equation}
and $\sigma(V_{\beta})=\sigma(V+\beta\idE)$.
One may also write the operator $f(V)$ in the equivalent way:

\begin{equation}\label{fofv}
 f(V) = \sum_{\lm\in\sigma(V_{\beta})}f(\lm)\ket{\lm}\bra{\lm}.
\end{equation}
Thus, in view of the Eq.~(\ref{eigen}) and~(\ref{fofv}) the  
$X$ takes the final form

\begin{equation}\label{sol}
 X = \sum_{n}x_n\ket{\phi_n}\bra{\phi_n},
\end{equation}
where  abbreviation $x_n\equiv f(V_n+\beta)$ was introduced.
Note that  $x_n$ are the eigenvalues of the operator $X$, that is $X\ket{\phi_n}=x_n\ket{\phi_n}$.
The operator above is Hermitian, \ie,  $X=X^{\dagger}$. Since for every
parameter $\alpha$ and $\beta$ the function $f$ specified by the Eq.~(\ref{fun})
takes positive values, thus the eigenvalues $x_n$ are positive. Therefore, the
solution we obtained in the Eq.~(\ref{sol}) is positively defined operator.
Moreover, as was pointed out in~\cite{gardas2} there exists at least one more solution
of the Eq.~(\ref{sricc}), in the case where $H_{E}$ and $V$ were chosen to be the
ones that describe the spin-bath (see Eq.~(\ref{spin1})).  This  situation  also  occurs
when the operators $H_{E}$ and $V$ have more general form, then the second solution is
given by 

\begin{equation}\label{sol2}
  \bar{f}(\lambda)= \frac{-\lambda-\sqrt{\lambda^2+\al^2}}{\al},
  \quad \lambda\in\sigma(V_{\beta}).
\end{equation}
It can be easily verified that this function determins the negatively defined operator, that
is to say

\begin{equation}
 \bar{X} = \sum_{n}\bar{x}_n\ket{\phi_n}\bra{\phi_n},
\end{equation}
and $\bar{x}_n:=\bar{f}(V_n+\beta)$. Observe that $\bar{X}=\bar{X}^{\dagger}$, \ie, this 
operator is also Hermitian. Note also that $f(\lambda)\bar{f}(\lambda)=-1$, for
$\lambda\in\sigma(V_{\beta})$. This may be verified directly, or may also be thought of as a consequence
of the Viet'a formulas, if one thinks of $f$ and $\bar{f}$ as the solution of the quadratic
equation: $\alpha x^2+ \lambda x-\alpha=0$, for $\lambda\in\sigma(V_{\beta})$.
We want to emphasize however, that the Riccati Eq.~(\ref{sricc}) is \emph{not} a simple
quadratic polynomial, even though it may appear so. As a result there might exist other
solutions of this equation that are not specified by the well-known formula for roots of
the quadratic equation. 

At this point natural questions may by asked, for instance which operator, $X$ or $\bar{X}$
(or any other, if it exist) should by used to diagonalize the block operator matrix $H_{QE}$?
What difference (if any) does it make? Of course, each solution of the Riccati equation will
diagonalize the operator matrix with which it is associated. Nevertheless, in certain cases 
it may by convenient to chose one solution instead of the other. For example by studying limiting 
cases like $\alpha\to 0$. Indeed, if $\beta\not=0$, then we find that
(recall that $f\sim1/\bar{f}$)

\begin{equation}\label{limit}
 \lim_{\alpha\to 0} f = 0,
 \quad
 \lim_{\alpha\to 0}\bar{f}=-\infty.  
\end{equation}
Therefore, $X\to 0_{E}$ while $\bar{X}\to -\infty$ as $\alpha$ goes to $0$. This means that the first
solution is a continuous function of the parameter $\alpha$, including the $\alpha=0$ value even though the second operator, \ie, $\bar{X}$ does \emph{not} exist in that point at all. 
As a result, if in the process of analysis one decides to use the second solution $\bar{X}$ then one might meet serious
problems taking the limits $\alpha\to 0$. Furthermore, as was mentioned earlier the first solution is 
a positively defined operator, thus it is more suitable to manage. Henceforward, we will restrict further 
analysis to the operator $X$ given by the Eq.~(\ref{sol}) only.

\section{The exact reduced dynamics}\label{dynamics} 

We now use the solution above to construct the exact reduced dynamics of
the model described by $H_{QE}$. Obtaining the exact reduced dynamics of
the model under consideration, namely the one defined by the Eq.~(\ref{hamil}),
can be easily accomplished using the Eq.~(\ref{reduce}), as we mentioned
earlier. We begin with the construction of the evolution operator $U_t$ generated
by the Hamiltonian $H_{QE}$. Foremost, let us recall that the evolution operator
$U_t$ may be computed by applying the following formula

\begin{equation}\label{evolv}
 U_t = S_{X}\exp(-iH_dt)S_{X}^{-1},
\end{equation}
where $H_d$ stands for the diagonal form of the Hamiltonian $H_{QE}$ and in agreement
with Eq.~(\ref{diag1}) it takes the form

\begin{equation}\label{diag2}
H_d=
\begin{bmatrix}
 H_{\smal{+}}+\alpha X & 0_{E} \\
 0_{E} & H_{\smal{-}}-\alpha X
\end{bmatrix},
\end{equation}
where $X$ is given by~(\ref{sol}). Next let us observe that the inverse operator $S_{X}^{-1}$
is
 
\begin{equation}\label{invers}
  \begin{split}
S_{X}^{-1} &=
\begin{bmatrix}
 g(X) & 0_{E} \\
 0_{E} & g(X)
\end{bmatrix}
S_{X}^{\dagger} \\
&\equiv 
G(X) S_X^{\dagger},
  \end{split}
\end{equation}
where $g(X)$ is the function of $X$ and it is given by

\begin{equation}
g(\lambda)= \frac{1}{1+\lambda^2}, \quad \lambda\in\sigma(X).
\end{equation}
Note that $[X,V]=0$, thus $X$ and $V$ have the same eigenstate, namely $\ket{\phi_n}$.
Since $X=f(V_{\beta})$, the eigenvalues $x_n$ of $X$ are given by $x_n=f(V_n+\beta)$.
Obviously, $\sigma(X)=f(\sigma(V_{\beta}))$. The same arguments lead to the conclusion that
$g(X)\ket{\phi_n}=(g\circ f)(E_n+\beta)\ket{\phi_n}$, this implies that $\sigma(g(X))=(g\circ f)(\sigma(V_{\beta}))$. 
Using the Eqs.~(\ref{evolv})~-~(\ref{invers}) one may finally write the form of the
evolution operator of the total system, it reads

\begin{equation}\label{ut}
U_t =
G(X)
\begin{bmatrix}
 U_t^{+}+X^2U_t^{-} & (U_t^{+}-U_t^{-})X \\
 (U_t^{+}-U_t^{-})X & U_t^{-}+X^2U_t^{+}
\end{bmatrix},
\end{equation}
where $U_t^{\pm}:=\exp(-i(H_{\pm}\pm\alpha X)t)$.
Equation~(\ref{ut}) represents the evolution operator of the total system
$Q+E$ and it can be easily applied to any (not only to the factorable ones)
initial state $\rho_{QE}$ of that system, since it is written in $2\times2$
block operator matrix. At this point reduced dynamics $\rho_{Q}(t)$ may by
obtained, it has the following form

\begin{equation}\label{trace}
 \rho_{Q}(t)= \mbox{Tr}_{E}(U_t\rho_{QE}U_t^{\dagger}),
\end{equation}
where $\mbox{Tr}_{E}$ is the partial trace operation. Note that in general case
this may not be easy to accomplish, even though the Eq.~(\ref{trace}) might indicate
so. The reason for that is that to compute partial trace $\mbox{Tr}_{E}$ one       
needs to write operator $\rho_{QE}(t)$ in the $2\times2$ block operator form.  
   
\section{Kraus representation} \label{kraus}
 
It is interesting to see how the results of the previous section are
related to the standard description of the completely positive map via
so called Kraus sum representation:

\begin{equation}\label{sum}
  \rho_{Q}(t)=\sum_{\mu} K_{\mu}(t)\rho_{Q} K_{\mu}(t)^{\dagger},
\end{equation}
where the Kraus matrices $K_{\mu}(t)$ satisfy following completeness relation
$\sum_{\mu}K_{\mu}(t)K_{\mu}(t)^{\dagger}=\idQ$. It is well established that it is
possible to derive the Eq.~(\ref{sum}) from the Eq.~(\ref{trace}) assuming that no 
correlations between the system and its environment are present initially
~\cite{alicki,geometry}. The generalization to the case when initial
state is not factorable is also possible~\cite{korelacje,*erratum,KrausRep} 
(see discussion below). Nevertheless, in practice finding the Kraus matrices is
impossible  in most cases. We will show how to construct those matrices for the
system we study. To accomplish this goal let us rewrite $S_{X}$ in the following manner

\begin{equation}\label{sf}
  \begin{split}
S_{X} &= \sum_{n}
\begin{pmatrix}
 1 & -x_n \\
 x_n & 1
\end{pmatrix}
\otimes
\ket{\phi_n}\bra{\phi_n} \\
&\equiv
\sum_nF_n\otimes
\ket{\phi_n}\bra{\phi_n}.
   \end{split}
\end{equation}
We also used resolution of the identity $\idE$ in the 
$\ket{\phi_n}$ basis, that is to say $\idE=\sum_n\ket{\phi_n}\bra{\phi_n}$.
Note the inverse operator $S_{X}^{-1}$ can by written in a similar fashion, namely
 
\begin{equation}\label{isf}
S_{X}^{-1}=\sum_nF_n^{-1}\otimes\ket{\phi_n}\bra{\phi_n},
\end{equation}  
where $F_n^{-1}$ is the inverse matrix of $F_n$. Since we have $\mbox{det}(F_n)=1+x_n^2>0$
it always exists. Due to the fact that  $F_n^{-1}=F_n^{\dagger}/\mbox{det}(F_n)$ holds we can rescale $F_n$ namely $F_n\rightarrow 1/\sqrt{\mbox{det}(F_n)}F_n$ so it becomes the unitary operator. Furthermore, the
Hamiltonian~(\ref{diag2}) may be rewritten as

\begin{equation}\label{diag3}
  \begin{split}
   H_d & = \sum_n
   \begin{pmatrix}
    h_n^{\smal{+}} & 0 \\
    0 &  h_n^{\smal{-}}
   \end{pmatrix}
   \otimes
   \ket{\phi_n}\bra{\phi_n} \\
   & \equiv
   \sum_nH_n^d
   \otimes
   \ket{\phi_n}\bra{\phi_n},
  \end{split}
\end{equation}  
where $h_n^{\smal{\pm}}:=(E_n^{\smal{\pm}}\pm\beta)\pm\alpha x_n$ and
$E_n^{\smal{\pm}}:=E_n\pm V_n$ are the eigenvalues of $H_{\smal{\pm}}$ 
(note $\sigma(H_{\smal{\pm}})=\sigma(H_E\pm V))$). Combining Eq.~(\ref{evolv})
and Eqs.~(\ref{sf})~-~(\ref{diag3}) we obtain

\begin{equation}\label{evolv2}
 U_t=\sum_nU_n(t)\otimes\ket{\phi_n}\bra{\phi_n},
\end{equation}
with the unitary matrices $U_n(t)=\exp(-iH_nt)$ and $H_n=F_n^{-1}H_n^dF_n$.
One can see from the form of the evolution operator above and Eq.~(\ref{trace})
that we finally have

\begin{equation}\label{kraus2}
  \begin{split}
 \rho_{Q}(t)&=\sum_n\rho_n U_n(t)\rho_QU_n(t)^{\dagger} \\
            &\equiv
         \sum_n K_n(t)\rho_QK_n(t)^{\dagger}.
  \end{split}
\end{equation}
In the Eq.~(\ref{kraus2}) $\rho_n:=\bra{\phi_n}\rho_E\ket{\phi_n}$ and the Kraus matrices are 
defined as $K_n(t):=\sqrt{\rho_n}U_n(t)$. Therefore, the operator sum
representation of the model we study is found.
 
 \subsection{Connection with the Riccati diagonalization}
   \label{sub:Rdiag}
Interestingly, the explicit form of the matrix $H_n$ reads

\begin{equation}
 H_n =
  \begin{pmatrix}
  E_n^{\smal{+}}+\beta & \alpha \\
  \alpha &   E_n^{\smal{-}}-\beta
  \end{pmatrix},
\end{equation}
and it does \emph{not} depend on the eigenvalues $x_n$ of the operator $X$. Therefore, one may
draw the conclusion that the Kraus matrices obtained in Eq.~(\ref{kraus2}) do not depend on $x_n$.
To solve this puzzle, notice first that in order to compute the Krause matrices one needs to determine the
``evolution'' operator $U_n(t)$. The later requires diagonalization of its ``generator'' $H_n$.
Because $H_n=F_n^{-1}H_n^dF_n$ the dependence of the Kraus matrices on the eigenvalues $x_n$ is 
``hidden'' in a way, in the similarity matrix $F_n$. It is 
important to realize that the diagonalization procedure $H_n\xrightarrow{F_{n}} H_n^d$ differs from 
the standard diagonalization routine, which is based on the eigenvalue problem for the operator $H_n$.
This new kind of algorithm is called \emph{Riccati diagonalization} and was recently introduced
in~\cite{Rdiag}. One may see that it arises naturally
in our analysis. Let us also recall that the similarity matrix $F_n$ is composed of the $n$th eigenvalue
$x_n$ of the operator $X$, which is the solution of the Riccati Eq.~(\ref{sricc}), while the eigenvalues 
of the matrix $H_n$ are the solution of the following characteristic equation associated with $H_n$: 
 
\begin{equation}\label{char}
 \lambda^2-\lambda\mbox{Tr}(H_n)+\mbox{det}(H_n)=0,
\end{equation}
We have already found its solution indirectly in the Eq.~(\ref{diag3}). The roots of this equation are given
by $\lambda^{\smal{\pm}}_n=h_n^{\smal{\pm}}$. The corresponding eigenvector (not normalized) reads

\begin{equation}\label{vectors}
\boldsymbol{\lambda}^{\smal{+}}_n
  = 
 \begin{pmatrix}
  -\bar{x}_n \\
     1
 \end{pmatrix},
\quad
\boldsymbol{\lambda}^{\smal{-}}_n
  = 
 \begin{pmatrix} 
  -x_n \\
    1
 \end{pmatrix}.
\end{equation}
Therefore, the similarity (not unitary) matrix $G_n = (\boldsymbol{\lambda}^{+}_n,\boldsymbol{\lambda}^{-}_n)$
the one that also diagonalizes $H_n$ takes the form

\begin{equation}
  G_n =
  \begin{pmatrix}
   -\bar{x}_n & -x_n \\
    1  &  1
  \end{pmatrix}.
\end{equation}
The connection between similarity matrices $F_n$ and $G_n$ is following. 
If $F_n\equiv (\boldsymbol{\xi}_n^{\smal{1}},\boldsymbol{\xi}_n^{\smal{2}})$, then
$\boldsymbol{\xi}_n^{\smal{1}} =x_n\cdot\boldsymbol{\lambda}_n^{\smal{+}}$ and
$\boldsymbol{\xi}_n^{\smal{2}} =\boldsymbol{\lambda}_n^{\smal{-}}$, 
 since $x_n\bar{x}_n=-1$. Therefore the matrix $F_n$ is also composed of the eigenvectors of
$H_n$, yet    the  matrices  $G_n$  and  $F_n$  are  \emph{not}  similar,  \ie,  the invertible matrix $P_n$ such that $G_n=P_nF_nP_n^{-1}$ does not exist.
It immediately follows from the  
fact that $\mbox{Tr}(G_n)\not=\mbox{Tr}(F_n)$ as well as $\mbox{det}(G_n)\not=\mbox{det}(F_n)$.
Since the similarity transformation preserves the trace and determinant, thus the matrices $G_n$
and $F_n$ can not be similar.
From the consideration above one may easily grasp the main difference between the two methods.
In the standard approach one needs to determine the eigenvalues $\lambda^{\smal{\pm}}_n$ of $H_n$
as was mentioned above (\ie, one need to solve Eq.~(\ref{char})).
On the other hand to use the Riccati diagonalization schema to our advantage we need to find $x_n$
(\ie,  resolve the Riccati Eq.~(\ref{sricc})). Note that in the case of  the ordinary matrix,
the Eq.~(\ref{sricc}) becomes the quadratic equation, but it differs from the characteristic Eq.~(\ref{char}). 
In our model the relation between $\lambda^{\smal{\pm}}_n$  and $x_n$ may be summarized as

\begin{equation}\label{diff}
\lambda^{\smal{\pm}}_n = (E_n^{\smal{\pm}}\pm\beta)\pm\alpha x_n.
\end{equation}

 \subsection{Initial correlation}\label{sub:corr}

Let us notice that if the initial correlations between system of interest and its
environment are present, \ie, $\rho_{QE}$ takes the form

\begin{equation}\label{corr}
\rho_{QE}=\sum_{ij}\gamma_{ij}\rho_{Q}^{i}\otimes\rho_{E}^{j},
\end{equation}
for some \emph{not} factorable complex number $\gamma_{ij}$, namely 
$\gamma_{ij}\not=\gamma_1^i\gamma_2^j$, then reduced dynamics~(\ref{trace}) 
\emph{cannot} be written in the form~(\ref{sum}). This not so obvious, since
there are cases in which, even though initial correlations are present the reduced
dynamics can still be written in the Kraus form~\cite{OhKwek}. 

Yet, for a finite dimensional environment there exists a simple criterion~\cite{KrausRep}. 
It allows one to verify when the given state $\rho_{Q}(t)$ possesses the operator sum 
representation~(\ref{sum}) if initial correlations are present. The necessary and 
sufficient condition for the later to holds true for \emph{any} initially correlated state is that
the joint dynamics has to be locally unitary, \ie, the evolution operator $U_t$ for the total system
needs to be of the form

\begin{equation}\label{local}
U_t = U_{Q}(t)\otimes U_{E}(t).
\end{equation}
The operator $U_{Q}(t)$ $(U_{E}(t))$ describes the evolution of the system $Q$ $(E)$ alone.
Observe that if the evolution of the total system is \emph{not} locally unitary that this
not necessary need to implies that $\rho_{Q}(t)$ does not posses Kraus representation for 
\emph{particular} initially correlated state $\rho_Q$.
From the Eq.~(\ref{evolv2}) one can readily see that the evolution operator~(\ref{ut})
does not have the form~(\ref{local}) as one may expected. Note that from~(\ref{corr}) and 
Eq.~(\ref{evolv2}) we obtain

\begin{equation}
 \rho_Q(t) =\sum_{n}\sum_{ij}\varepsilon_{ij}^nU_{n}(t)\rho_Q^iU_{n}(t)^{\dagger},
\end{equation}
where $\varepsilon_{ij}^n =\gamma_{ij}\bra{\phi_n}\rho_E^j\ket{\phi_n}$.
Therefore, even if in this general case the reduced dynamics can not be written 
in the operator sum representation, one can still describe the evolution of the system
in a manageable way.

\section{Summary}

In this paper we investigated the Riccati algebraic equation  associated  with  the
Hamiltonian defining the process of decoherence in the case of one qubit. It was shown
that if the environment  is  $\tau-$symmetric, where $\tau$ is  antilinear involution,
then $\tau$ is  the  solution  of  the  Riccati  equation under consideration. 
We indicated that even though the solution of the  Riccati equation has been
found it can not be applied to obtain the reduced  dynamics,  due to the problem with
standard procedure allowing one to solve Schr\"{o}dinger equation. We wish to emphasize
that this result does \emph{not} contradict with the previous  paper~\cite{gardas}, where
we claim that the solution of the Riccati equation enables one to rewrite the evolution operator
generated by the Hamiltonian $H_{QE}$ as $2\times 2$ block operator matrix. Of course, 
the reason that problems occur is that the solution of the Riccati Eq. is antilinear. 

Furthermore, we provided a full resolution of the problems introduced in~\cite{gardas}
for the case when operators defining the environment commute with each other (regardless of the
existence of any symmetry in the system). We also derived the operator sum representation for
that model assuming no correlations between the systems are present initially. Moreover, we
also showed how to obtain the solution if the initial state of the total system in not
factorable.
This result is a direct generalization of the system discussed in~\cite{gardas2}. We
also indicated that the recently derived schema of so called Riccati diagonalization
arises naturally in the model we considered.

\begin{acknowledgments}
We acknowledge the financial support by the Polish Ministry of Science
and Higher Education under the grant number N N519 442339. The author
would like to thank Jaros\l{}aw Adam Miszczak for helpful comments and
suggestions. 
\end{acknowledgments}


	 
%

\end{document}